# A Circular Trajectory of Fixed Points around a Virtual Rotation Axis for Computed Tomographic Image Reconstructions

Kyungtaek Jun[a,*], Seokhwan Yoon[b], Kyu Kwon[a]

[a]Seoul Hayanara Research Center, Gyeongsangnam-do, Changwon-si 51495 Korea

[b]Department of Dentistry, Seoul National University, Seoul 03080 Korea

**Abstract**

Since X-ray tomography is now widely and usefully adopted in many different areas, it becomes more crucial to find a robust routine of handling tomographic data to get quality reconstructed images. Though there are existing techniques such as center of mass calculation, image registration, and reconstruction evaluation, with their own limitations, it seems helpful to have a more automated method to remove the possible errors that hinder clearer image reconstruction. In this paper, we proposed an alternative method and new algorithm using the sinogram and the fixed point. A new physical concept of Center of Attenuation (CA) was also introduced to figure out how this fixed point is applied to the image reconstruction with errors we further categorized. Our technique showed a promising performance in restoring images with translation and vertical tilt errors. The limitation of our method is discussed and the area that future researches are needed is also mentioned.

*Corresponding author

E-mail address: ktfriends@gmail.com (Kyungtaek Jun)

# 1. Introduction

X-ray tomography is considered more than essential nowadays in various fields[1], and combined with advanced X-ray sources, it provides us with more sophisticated scientific insights[2]. As Wang et al.[3] once argued, it is desirable to achieve an automated data processing with minimum human interaction due to the demand of high-throughput experiments. Given the fact that we only get the 2D projections from the x-ray penetration to form the internal 3D structure of a sample, a number of scientists have shared their insights to get rid of impairing effects in getting a quality reconstruction, translation and tilting errors, for example. They have focused on the determination of the center of rotation (CoR), and there are mainly three families of methods. The first category uses pairs of projection images taken from the reverse viewing angles (at 0-180 degrees)[4]. They perform the image registration of these pairs to calculate the offset of CoR. This method is often considered efficient, however, Vo et al.[5] argued that it is not feasible especially when the projections have low contrasts or the optics system have fixed defects. The second method evaluates the projection image from the reconstruction using a parameter to measure the quality of image reconstruction and to calibrate the relative offset of the rotation axis[6]. This technique is widely used and has its own strength in terms of using all the available information, but it's often time-consuming and inapplicable for the reconstruction with artifacts[3]. The last one considered the center-of-mass (CM)[7]. To make it work, the sample should be within the field of view, which is not always possible.

Even with these numerous efforts to minimize and correct the systemic defects, not all the errors are perfectly corrected and often the process is too time consuming and laborious. Also, most of the error correction was done only for small amount of error. Hence it is always useful to figure out a new method and algorithm to correct the flawed information to obtain better quality image reconstruction.

In this report we propose an alternative approach to correct those errors, mainly translational and tilting errors, using the sinogram and the fixed point. Scientist also have used a similar approach of using sinogram (Vo et al.), however, ours is different in a way that we are more focused on the layer of a sinogram. As a sinogram is an accumulated projection image, each column of a sinogram indicates an angle of a projection row of an object. If an image is tilted, which happens quite a lot in the nanometer scale projection where vibration of a system is not scarce, the layer that is tilted causes an inevitable error for the reconstruction. Besides, we did not necessarily try to determine the center of rotation. Instead, we focused on the fact that there must be a "not-moving" point in the projected image of a specimen just like the concept of center-of-mass. The specimen for the BNL tomography is picked up and replaced for the energy level checkups, at every 60 projections among the 1200s for the entire 180 degrees angle rotation. It is hardly possible to determine the CoR of the specimen because, even though the RA hasn't changed, it looks like changed when the specimen itself was rotated. We did not seek to determine the CoR; instead, we aligned the vertical center line of projection to the CoR to make it like a virtual RA, which was enabled by our algorithm.

## 2. 2D Image Reconstruction from a Projection Set

### 2.1. The relative position of an object in the real space and the sinograms

A sinogram involves the information about a specific layer of an object and accumulates the projection shadows taken from each angle to build a reconstruction image of the layer. This sinogram will later be transformed into a reconstructed image through the inverse Radon transform. Apparently, an errorless reconstructed image will be obtained only with an ideal sinogram or an ideal projection set, and a necessary condition for the ideal sinogram is that its pattern is changed sequentially following the projection angles.

When the location of an object is changed on the stage, the sinogram pattern will be also changed while the projection shadow from an angle remains the same. (1)

There might be slight differences in the projection shadow due to the limit of image digitalizing techniques; this can be approximated by a linear interpolation.

Where a projection shadow lies on the sinogram is closely related to where an object lies in the real space; this relationship should be first investigated before we further use sinograms for our analysis. The object on the rotating stage is known to rotate around the rotation axis (RA) and it means that any single point of the object has its own circular trajectory, which will be expressed as a sinusoidal function in the sinogram. The point on the RA will be the straight line across the center of the sinogram. Our hypothesis is that if we get the projection set of a specific object and if we know the ideal sinogram pattern of this object, we can modify the real sinogram pattern to meet the ideal one and consequently get the errorless reconstructed image.

We brought the projection image of a simple cylinder to show how a sinogram changes accordingly as a specimen moves on the stage. The sinogram was constructed by a projection image of the circular specimen, a cross-sectional image of a cylinder in 2D. Fig. 1(a) shows when the specimen was located on the center of the stage. The stage rotated clockwise for the entire 180° or 360°, and the x-ray beam was fixed at $\theta = 0$. The sinogram pattern is linear in this case. One can easily spot the changes of the sinogram pattern when the cylinder was moved toward the parallel beam on the stage in Fig. 1(b), and when it was moved to the left from the beam in Fig. 1(c). The projection shadow from an angle remains the same, even when the specimen was moved to a different spot and had a different sinogram pattern. If we can somehow modify the sinograms in Fig. 1(b) or 1(c) to make them just like the sinogram of Fig. 1(a), it has the same effect as physically moving the object from where it was to the center on the stage, while the recorded image is preserved. This saves the present efforts of determining the COR and setting the specimen on the RA, for the simple modification of sinogram will yield the same result.

## 2.2. Continuity & Discontinuity by an Object movement and a CT system Vibration

Assumed that the projection images are all clear, it is the translation error that we can first think of in the 2D space during the beam time. We categorized translation errors into two; the orthogonal translation and the parallel translation, which are the two elements in the basis that we decided to consider in this study. By the orthogonal translation, we mean that the specimen is moved vertically from the parallel beam at the projection angle $\theta$, while the parallel translation means that the specimen is moved toward to or away from the parallel beam. These two types of errors let us distinguish whether a certain error can be corrected or not, because the error occurring horizontally to the beam is hard to notice, and therefore hard to calculate the distance of shift. And significantly, the pattern of sinogram flows continuously showing no discontinuity in parallel translation (Fig. 2(a)).

The error occurring vertically to the beam shows a definite cut-off point as the graph of Fig. 2(b) illustrates. In Fig. 2(b), the error appeared when the parallel beam was shoot perpendicularly to the stage and the cut-off of the sinogram was reflected at the 90 degrees of $\theta$. The error that arises when the specimen is moved perpendicularly to the beam is always reflected identically on the projection and the discontinuity of equal amount is also found in the sinogram. As a result, we can spot the exact point where the discontinuity happens and correct this error mathematically.

If we have an orthogonal translation error in the *e*-th column, the correction process has to start from the point where the error occurred and the rest of sinogram has to be modified accordingly. Let us calculate how far the shadow has to be compensated in order to make this inaccurate sinogram an errorless, ideal one. $D_n$ is the distance to which the $n$-th column has to be moved in the sinogram, and it is written as follows:

$$D_n = d * \cos(\theta_n - \theta_e), n \geq e \text{ or } n < e$$

$\theta_n$ indicates the angle of $n$-th column and $\theta_e$ is the angle at which the orthogonal translation error occurred in the sinogram. $d$ is the distance that the object is vertically moved to, and can be measured through the two columns $\theta_{e-1}$ and $\theta_e$ in the sinogram, for the sinogram exactly reflects the shift. To find $d$, we used a certain section of $e$-th column and the section will be compared to the part of $(e-1)$-th column. Here, calculating $d$ is an issue of the optimization problem and $f_e(t)$ has an absolute minimum at $t = d$ and $n = e$.

$$f_n(t) = \frac{1}{m}\left\{\sum_{k=1}^{m}|p_{n-1}(k+s-1) - p_n(k+s-1+t)|\right\}, -s < t < s$$

A total pixel number $m$ is used in calculation. $s$ is the starting point in the $(e-1)$-th column. $p_i(j)$ indicates the value of a pixel position at $i$-th column and $j$-th row in sinogram; $p_1(1)$ indicates the initial pixel point. If the object's shadow length in the sinogram is small enough compared to the projection size, $m = \frac{1}{2} * ny, s = \frac{ny}{4}$ is enough to use ($ny$ is the total vertical pixel number). $t$ is an integer and indicates translation. To simplify the calculation, we assumed that the pixel length is the unit length. To obtain the real value of $d$, it is possible to use the symmetric axis of a quadratic equation with $d - 1$, $d$, $d + 1$ and their $f_n$.

For example, the values of $d = 49$, and $\theta_e = 90°$ were obtained when the sinogram of Fig. 2(b) was analyzed. Any abnormal peaks were checked to figure out $\theta_e$ in the graph where x-axis indicates the angle and y-axis indicates the average difference value $d$, which is the distance between two columns of $t = 0$. After finding $\theta_e$, we calculated d by the minimum value of $t$, $0 \leq t \leq 60$. Using the vertical movement value driven from the following formula, $D_n = 49 * \cos(\theta_n - 90°) (\theta_n \geq 90°)$, the sinogram of Fig.2(b) was transformed to an ideal one shown in Fig.1(a). Generally, it will be transformed to an similar-to-ideal one which satisfies the optimization formula, because we don't exactly know the column information of the ideal sinogram at $\theta_e$. Fig. 3 shows how we can apply the orthogonal translation algorithm to the real projections.

## 2.3. Correction Algorithm

In the real settings of computed tomography (CT) system, the translational error is not only either a horizontal or vertical error but also a combination of both. Thus, a solution for the orthogonal translational error is not general enough. As we mentioned earlier, the horizontal shift of an object is hard to detect on the projection image and thereby, hard to calculate. This leads us to a need to approach this issue using a fairly new concept and a point of view. Knowing that the errors in the real settings are complicated and often a mixture of different kinds, one should start with understanding and defining the relationship between the real space and the sinogram, which is actually a projected and reconstructed version of real space, so as to further use it to draw a general and automated method of correction.

Ideally, the centerline of a sinogram is made up of the projection of RA. Any single point in the real space will move on a circular trajectory rotating around the RA and this will correspond to a function on the sinogram. Let's say there is a point $p$ and it moves on a circular trajectory, this point $p$ will be transformed into a curve on the sinogram and represented by the sinusoidal function $T_{r,\varphi}$.
The circular trajectory of a point $p$ in the real space corresponds to a curve drawn by the sinusoidal function in the sinogram. (2)

$$T_{r,\varphi}(\theta) = r * \cos(\theta - \varphi), 0 \leq \theta < 180°$$

We defined $r$ is the distance between the rotation axis and the point $p$. $\theta$ is the projecting angle, and $\varphi$ is the angle between the line $\overleftrightarrow{Op}$ and the orthogonal line to the projection angle at $\theta = 0$ (Fig. 4). $O$ is the RA.

Using (1) and (2) which were mentioned earlier in our study, we can alter an inaccurate sinogram to a correct one. $T_{r,\varphi}$ is a function that shows how a specific point $p$ in the real space moves on the sinogram. It means that we can keep tracing a point of the real space on the sinogram and further use it as we intend to. Being the center of a circle, the point $p$ on the sinogram in Fig. 4 will be always on the center of the projection shadows. It follows the curve marked as a solid line on the sinogram since the specimen itself is off the center (see the sinogram on the right side of Fig. 4). The point $p$ is a center-of-mass for a circular object if the object has an identical medium and acts as a fixed point to represent the same spot even when the projection angle changes. With the point $p$ translated on the function $T_{0,\varphi}$, that is to say, with each column in the sinogram moved so that the point $p$ is on the center and the solid lined curve is now laid linearly on the center line of the sinogram, the center of an object will match exactly to the center of projection and the sinogram will become linear. When applied to a real space, this virtual translation has the same effect as we physically set the center of an object on the RA.

Nevertheless, real objects are not always cylindrical. They rather come in much more complicated structures, and as a result, it is not an easy issue to define a fixed point on the projections. We came up with a new physical concept called the center of attenuation (CA) to make this issue simpler. The CA is a similar idea with the center-of-mass, but is different in a way that a unit particle of an object is expressed not by the mass but by the x-ray attenuation and that it is calculated by the length.

When the relationship between the change of x-ray intensity on a logarithmic scale and the specimen length is not linear after the x-ray beam penetrates the specimen, it should be first transformed into something with linearity and utilized to figure out the CA.          (3)

The value of attenuation should be changed into the one expressed in length to calculate the CA, and when it is changed, it is called the modified attenuation (MA).

When an x-ray beam is shot to an object, the object absorbs certain amount of the energy and the rest of attenuated energy arrives on the detector. Then we get the projection image. As any object has a center-of-mass, we assumed that any object has an invariable center of attenuation, and it is fixed on a certain spot, either inside or outside of the specimen, acting as a fixed point that does not change depending on the projection angle. This fixed point can be calculated from the projection images obtained from each angle of x-ray penetration. It will lie on the sinogram satisfying the $T_{r,\varphi}$ function that was mentioned earlier and this is how we get an ideal projection set.

The calculation for CA is very similar with the one for center-of-mass.

First, a virtual rectangular coordinate and unit cubes are adopted in the real space and each vertex is the whole number on the coordinate. Let's assume that the specimen placed in the coordinate is composed of $n$ unit cubes. The $i$-th cube among the n cubes of the object has a certain modified attenuation on its designated location and it is represented as $a_i$, which can be linearly calculated through a series of calculation processes.

Here is how we mathematically identify the location of CA and $\overrightarrow{CA}$ on the coordinate we just assumed.

$$\overrightarrow{CA} = \frac{1}{A}\sum_{i=1}^{n} a_i \vec{r_i}$$

$A$ is the total sum of $a_i$, and $n$ is the total grid number of the object in the real space. $\vec{r_i}$ indicates the center of $i$-th grid. $\overrightarrow{CA}$ is one of the fixed points in the real space and particularly significant in that it will be always projected and we can trace it later on the projections. So, $\overrightarrow{P_{CA}}$, a projected position of $\overrightarrow{CA}$, can be calculated in the projection image.

$$\overrightarrow{P_{CA}} = \frac{1}{A}\sum_{i,j} p_{ij}\overrightarrow{r_{ij}}$$

Here we see that $A = \sum p_{ij} = \sum_{i=1}^{n} a_i$. $p_{ij}$ is the $(i,j)$ pixel value of the projected specimen in the 2-dimensional projection and $\overrightarrow{r_{ij}}$ is the center of the pixel. As the $a_i$ is expressed by length, the $p_{ij}$ also should be linearly expressed.

In the projection image, the attenuation value of the area except for the computational domain for the object shadow should be zero in the ideal status. If it is not zero, $\overrightarrow{P_{CA}}$ might not be able to act as the fixed point. So we need to modify the attenuation value of this area to have at least the average of zero when it is not exactly zero, ensuring the same amount of MA is added or subtracted to the $p_{ij}$ area.

$\overrightarrow{CA}$ in the real space becomes a specific point that we know and it is projected to be $\overrightarrow{P_{CA}}$ and expressed as the $T_{r,\varphi}$ function on the sinogram. The center of projection is always the center line of the sinogram. Therefore, when $\overrightarrow{P_{CA}}$ from each angle of x-ray beam is translated onto the center line $T_{0,\varphi}$ of the sinogram, it can be considered the same as the fixed point of an object is placed on the RA. In other words, we are able to get rid of all the translation errors in the 2-dimensional space by translating $\overrightarrow{P_{CA}}$ of each projection on the $T_{r,\varphi}$ function.

Fig. 5 shows how we applied this idea to the real images.

### 3. 3D Image Reconstruction from a Projection Image Set

There are only translation errors in the two dimensional projections. However, other kinds of error have to be taken into consideration when it comes to three dimensions. Mainly, three kinds of errors can be discussed; a translation error caused by the shift of a specimen, a tilting error caused when the RA is tilted, and the rotation error that is caused when a specimen spins on its own axis.

When we have a rotation error, some information of the image projected from certain angles is lost when the angles of x-ray beam and the rotation overlap. For instance, only the same face of an object is projected even though the beam supposedly takes every different projection, when the rotation of specimen itself countervails that of the beam. In this case, we are not able to obtain every projection image of every angle, namely a complete projection set, which is an essential element in x-ray tomography. And therefore, we will not discuss further on the issue in this study.

A translation error is the error that happens when the object is moved by any chance during the beam time, and the movement can be in three directions in the 3D space. A compensation for this error can be done through just the way we did with the translation error in 2D space, bringing $\overrightarrow{P_{CA}}$s on the $T_{r,\varphi,h}$ function.

$$T_{r,\varphi,h}(\theta) = (r * \cos(\theta - \varphi), h), 0 \leq \theta < 180°$$

The translation error in 3D is different from the one in 2D in that we should adjust the layer of CAs so that every CA is on the same layers. Hence, we first set a specific layer of projections as height $h$, and translate $\overrightarrow{P_{CA}}$s onto the $T_{r,\varphi,h}$ function of this layer. This is a significant procedure in the 3-dimensional reconstruction because it is simply not possible to reconstruct an image well using projections of different layers. To make our calculation easier, we will place the CAs on the RA ($r = 0$) of the projections' center layer ($h = 0$) and this process will be more beneficial if one considers using an optimization method to correct tilting errors. By doing this, we can compensate most of the translation errors and the CA is on the center of projection (Fig. 6).

### 3.1. Analysis of tilting Error in 3D

In 3-dimensional space, it is important to consider tilting errors because a tilted image carries information from different layers and consequently induces a flawed image reconstruction. Thus, if we compensate the tilting error, it means that we make one layer to carry all the information about a single part of an object.

To categorize and analyze tilting errors, we should further discuss about the object itself and the stage that it is placed on. If a cylindrical object on the stage is tilted and projected, we cannot tell whether it is the object or the RA of the stage that is tilted, only by looking at the projection image. However, those two should be distinguished and defined. When the RA is upright and only the object is tilted, it should not be called a tilting error and consequently, the projection set should not be corrected. Instead, it should be considered as we put some other object which originally has a tilted shape. As it sounds ironical, one would hardly get an ideal projection set if this kind of error is corrected because the original projection set was already correct. Nevertheless, it is hard to tell the difference between the tilt of an object and the one of a RA in the projection image; one should turn to the sinogram in this case because these errors will be seen more easily and clearly in the sinogram.

We can make sense of this logic better with the following figures. The Fig. 7(a) shows when the object is tilted whereas Fig. 7(b) shows when the RA is tilted. The sliced image of the specimen upon a same layer is identical in those cases. However, their sinograms show definitely different patterns.

In case of a tilted object, the sinograms of its top and bottom follow sinusoidal functions. When the RA is tilted, however, the heights are the same with the ones in the former case, while the patterns of the top and bottom go linear starting from each height. The sinograms obtained from the Fig. 7(a) are the ideal sinograms which will produce a correct reconstructed image; the others will not give us a correct reconstructed image since the layers are all mixed up during the projection. In short, we will not call the case of tilted object an error. It is because we can reconstruct a correct image without any correction procedures in this case. When it is the RA that is tilted, one should make sure that the tilting error is corrected.

In terms of the tilted RA, we can further classify two cases of leaning in detail; one is of a vertical direction and the other is of a horizontal direction toward the x-ray beam. The former is shown in the Fig. 7(b) and the latter is shown in the Fig. 7(c). In a vertical tilt, the polar angle increases when the azimuthal angle is either 90 or 270 degrees. On the other hands, the polar angle increases when an azimuthal angle is either 0 or 180 degrees in a parallel tilt. The azimuthal angle increases counter clockwise against the beam. The ideal RA which stands upright without any tilting error is assumed to be the reference for the polar angle.

It is the case of a vertical tilt when the RA is tilted against the axis of whole projection and the stage is rotating with the tilted RA. If we assume that the common layer is also tilted in parallel with the stage, namely tilted by the same angle in the projection, the projection information is now about a single layer, not mixed with the ones about other layers. Hence, if we only know how much the RA is tilted, and carry out a rotation compensation on the whole projection rotating around CA, we can figure out the common layer that we can concentrate on.

Contrarily, the parallel tilt carries information about more than just one common layer and it is hard to get a flawless reconstruction image in this case.

It is the projection image of a cylinder standing perpendicularly on the RA of the stage in the Fig. 8(a). Fig. 8(b) shows a projection image of the cylinder that is leaned in parallel with the beam and its depth of the shadow is changed, meaning the information of the image is changed. One can see the projection image of the same cylinder tilted vertically in Fig. 8(c). It is shown that the depth of shadow remains unchanged, even though the object is tilted; the projection information is not changed.

### 3.2. Correction algorithm for the vertical tilting of an object

In correction process, the priority is to know by which angle the RA is tilted. The whole projected images will be corrected accordingly to the tilted angle using the center line of whole projection, $T_{0,\varphi,h}$, as a fiducial line. To figure out the specific angle that the RA is tilted by, we used our CA in this study. Assuming that we virtually collected the every $\overrightarrow{P_{CA}}$ of each projection we get from a specimen and placed them on the charged coupled device (CCD), we thought that they would form a virtual line segment just like in Fig. 7(b). In fact, the line segment of $\overrightarrow{P_{CA}}$ stands for the trajectory of CA, and the RA is perpendicular to it. If this RA and $T_{0,\varphi,h}$ makes a certain angle, it is the angle by which we will rotate the projection set to meet the ideal one.

### 4. Discussion and Conclusion

In this study, we categorized the errors that possibly take place during the beam time. Although there certainly is a limit to what extent we can correct the reconstruction image in each case, we can modify most of the errors using our CA and get a close to ideal reconstruction image when it is the translation error. There are two cases for the tilting errors. When it is the vertical tilt, we get the ideal reconstructed image. Meanwhile, we will not be able to reconstruct a correct image with the projection set for the parallel tilt since the information of different layers are mixed up. Mostly, the tilting error is a mixture of both vertical and parallel tilts; the optimized projection set in which the vertical error is modified is the best deal in this case. For the rotation error, it seems impossible to reconstruct appropriately because we hardly get the complete projection set.

The method to figure out the angle of the vertical tilt is already discussed earlier. Even in a parallel tilt, the tilted angle can be sought with the same method. The angle of parallel tilt, α, satisfies $\sin\alpha = \frac{b}{a}$. Here, $a$ means the major axis and $b$ does the minor against the trajectory of CA. Whether the orientation of trajectory in the projection is clockwise or counter clockwise depends on whether the azimuthal angle of RA is 0 degree or 180 degrees. This method can be also applied to the mixed error of vertical and parallel tilts.

The CA we suggested in this study will function as the fixed point that is one of the intrinsic factors of an object. It works as an invariable point inside (sometimes outside)of an object which does not change even when the object is translated, or the RA is tilted. Nonetheless, we have to calculate this CA from the projected image so as to utilize it, which results in one limitation that the intensity variation of the beam reaching each cell of the CCD in the formula for $\overrightarrow{P_{CA}}$ should be linearly proportionate to the length of the specimen. That is, the attenuation value of each cell which consists of the image from the projection should have certain linear relationship with the length of the specimen, or at least should be changeable to have linear relationship. (see Supplementary Section 5). Particularly in soft X-ray tomography (SXT), there is a linear relationship[8] and a unique linear absorption coefficient measurement is given[9,10]. Scientists have successfully used full-rotation imaging of SXT for 3D images of various cell types with isotropic resolution[9,11,12,13,14,15,16]. They have tried to automate the tracking process of fiducial markers through a stack of projection images[17,18,19,20,21,22]. The trajectories of these fiducial markers follow $T_{r,\varphi,h}$ function; the fiducial markers can also be used as fixed points.

To obtain well reconstructed image, we should make sure to get a consistent x-ray density against the object regardless of projecting angles. In reality though, we sometimes get the different x-ray density according to the projecting angle changes, and then, the value of $A$ also should be changed in the formula applying the CA. It would be difficult to generally use CA for the reconstruction if the location of CA depends on the changes of x-ray density and the $A$ value. The changes of x-ray density (according to the angle) will appear in the sinogram in forms of the shadow intensity. Nonetheless, it won't affect the pattern of the sinogram. In Fig. 9, the pattern of ideal sinogram was maintained although there was a 50% decrease in x-ray density from the 90 degrees of $\theta$. The reconstructed image using the sinogram with the CA and the $T_{r,\varphi,h}$ function applied (Fig. 9(b)) showed no difference in terms of image itself when compared with the reconstructed image of Fig. 9(a), and the image of specimen in Fig. 9(b) was laid in the center. This proves to us that the location of CA does not vary despite the changes of $A$ value against each $\theta$ drawn by the different x-ray density. Also, we may be able to get a better image if we mathematically modify the x-ray density in order to even up the $A$ value.

The CA in our study is expected to significantly contribute to a better image reconstruction in x-ray tomography and to be utilized as a versatile tool. Nevertheless, there is a definite limit. For instance, when the specimen is mixed up with materials having greatly diverse ACs to the extent where it is out of linearity, it is hard to compensate the nonlinear area and this may result in errors. And it is not beneficial to use the CA in those cases when the specimen is projected with impermeable materials like metal to make it distinguished. It is better to project the specimen itself without any other distinguishing material when anyone wants to use CA in image reconstruction. The ring artifact due to the CCD defect could be another hurdle in CA application. We need to correct the ring artifact beforehand; the correction itself can also raise some changes in reconstruction errors.

When the specimen is longer than the CCD, its whole image is not projected and consequently, there is a limitation in applying the CA. We expect to discover another solution even in those cases if we can figure out the common layer on which the object is commonly projected and apply the CA on it.

In our study, we realized that every single point on the stage shows a periodic motion around the RA and the motion will be reflected as a sinusoidal function on the sinogram. If we try to optimize the errors, the optimization was only done to the moment of the error occurred in the sinogram, and the sinogram was vertically moved to connect the discontinuity without talking the sinusoidal function into account. This, unfortunately, leads to flawed image reconstruction. So, we had to use a formula like $D_n$ to modify our sinogram. This is shown quite intuitively in Fig. 10. An object was placed on the exact center-of-rotation, and the ideal sinogram of the object is found in the Fig. 10(a). If there is a vertical error at the 90 degrees, the sinogram will be like Fig. 10(b), and the error will be also seen in the reconstructed image. Simply moving vertically and linking the sinogram will bring us a sinogram like Fig. 10(c) and it will still yield incorrect, but different kind of reconstructed image. Our formula $D_n$, when the sinogram after the error is corrected, will bring a correct reconstructed image that is the same with the original image. Yet, it is not considered as a perfect reconstruction because what we used is the information of $\theta_{e-1}$, not the exact point of error, $\theta_e$, in optimization with the formula $D_n$. Therefore, we eventually need the function $T_{r,\varphi,h}$, using CA or FP.

In spite of the benefits of categorization, the errors often come about not as we categorized, but as a mixture of different kinds. It is not possible to perfectly restore the image even with the CA and the $T_{r,\varphi,h}$ function in those cases. The application will be much more extended if we can discover more than one fixed point besides CA and we can utilize each of them to the $T_{r,\varphi,h}$ function. Although further researches should be proceeded, we simply showed the example of this idea in Fig. 3. If there is an area whose boundary is definite and x-ray impermeability is comparatively high, it will be distinguished in the most of projections. We might be able to use its center as another fixed point. We searched for the trajectory of the fixed point in the sinogram of Fig. 3(b), and applied it to the function $T_{r,\varphi,h}$. It was transformed into the function $T_{0,\varphi,h}$ in Fig. 3(c).

Here, we see that it provided us with a clearer image when the $T_{0,\varphi,h}$ function was applied to the fixed point. It is expected most of the errors will be corrected when the fixed point and the CA applied in the $T_{r,\varphi,h}$ function in case of mixed errors.

Additional researches are further needed in the limits of applying the CA to real images. We will continue to seek for the correct algorithm in applying the CA and find out more traits that can be used as the fixed points. This endeavor will contribute to getting better reconstructed images even when the various errors coincide, when the whole image is not obtained because the specimen outlies the CCD, and when the parallel tilting error is included.

**Figures**

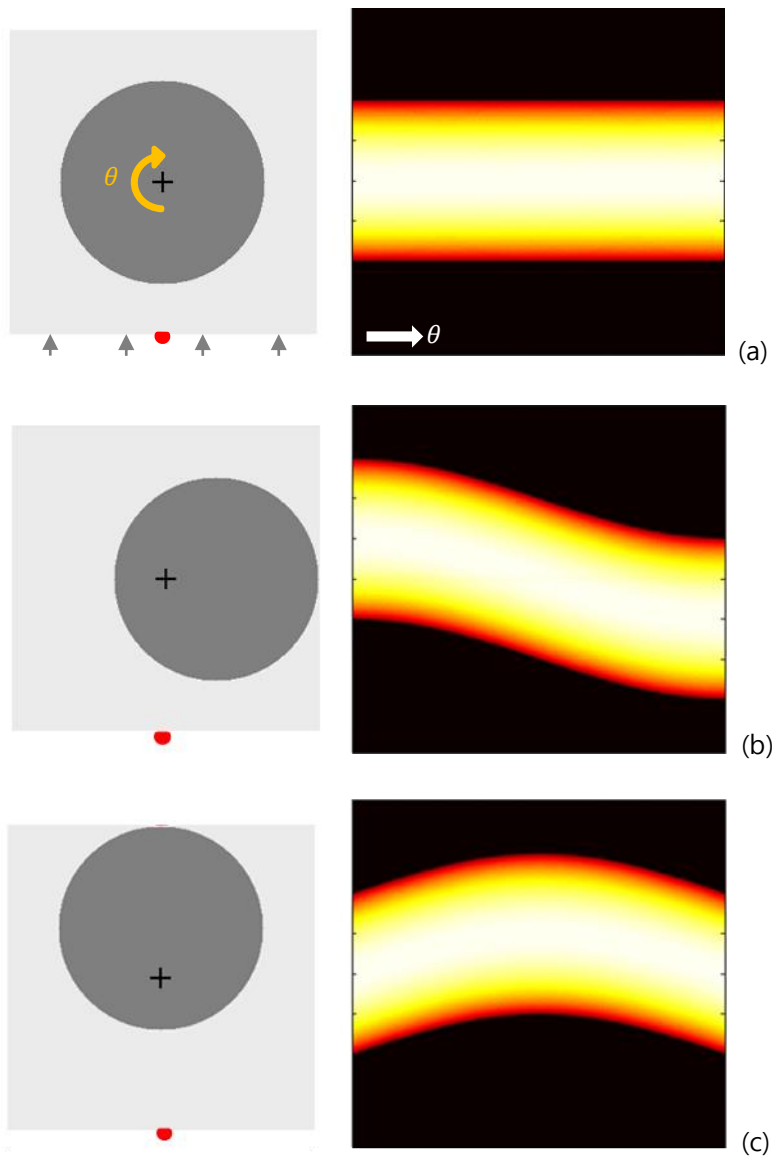

Fig. 1. The sinograms of specimen that were placed on several different part of the stage. Notice that we marked the stage with the red dot at the bottom to indicate $\theta$ is zero degree.

(a) The sinogram when the specimen rotates on the center of the stage

(b) The sinogram when the specimen is translated in parallel with the beam at $\theta = 0°$

(c) The sinogram when the specimen is translated vertically to the beam at $\theta = 0°$

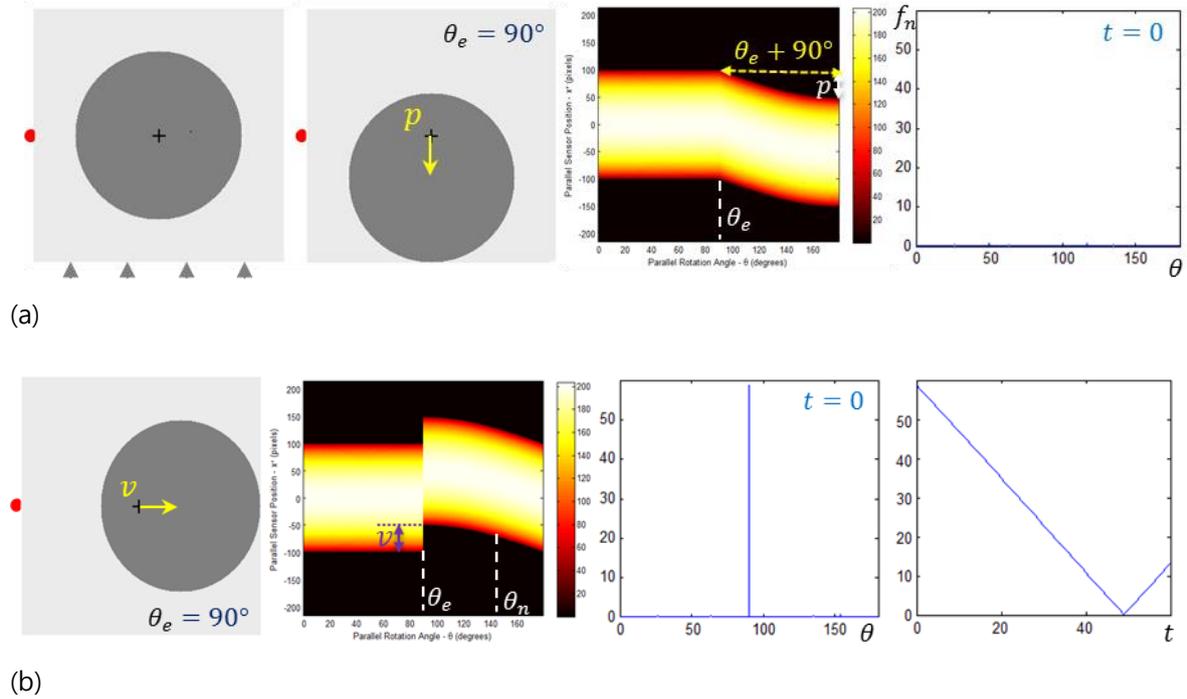

Fig. 2. The sinograms with the translation errors during the beam time.

(a) When the specimen is translated in parallel with the beam at $\theta = 90°$, the pattern of sinogram shows a change, but flows continuously. The graph which illustrates the value change of $f_n(t)$ to the angle, $\theta$, doesn't have any discontinuity, therefore, it is hard to spot the specimen.

(b) When the specimen is translated vertically to the beam at $\theta = 90°$, there is a discontinuity in the sinogram pattern. It is the same in case of the graph with the value change of $f_n(t)$ to the angle, $\theta$. The discontinuity is found at $\theta_e = 90°$, and the minimum is found at $t = 49$ in the graph

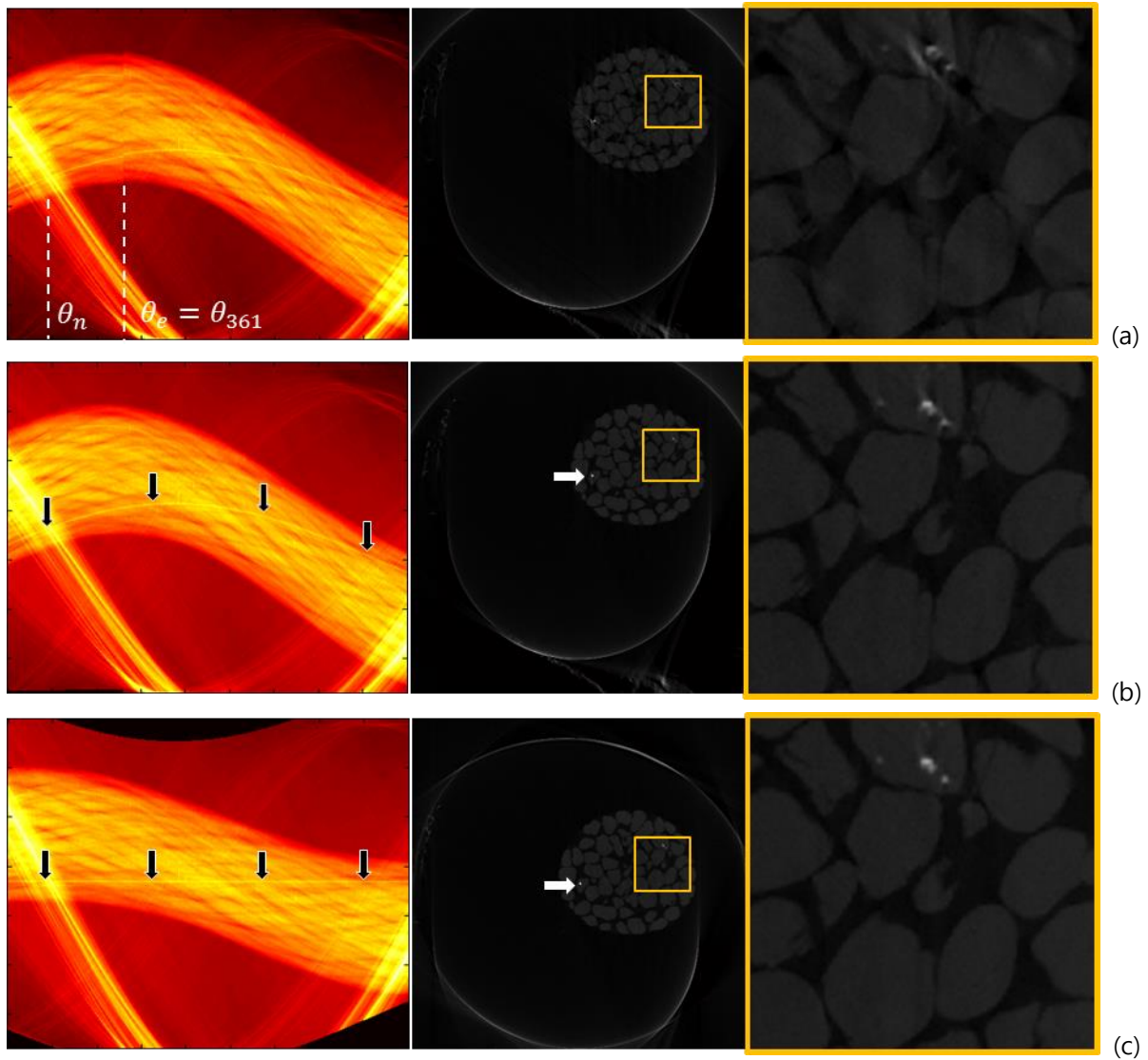

Fig. 3.The sinograms and reconstruction images of a layer of Hanford soil in a polyether ether ketone(PEEK) column from National Synchrotron Light Source (NSLS) X2B beamline at Brookhaven National Laboratory (BNL). The third pictures to the right are the magnified images of the white frame of the reconstruction images in the middle.

(a) Through the analysis of this sinogram, the specific $f_n(t)$ value was found at the discontinued spot; it was expected that a vertical translation error arose between the 360th and the 361th projections.

(b) We applied our formula $D_n$ to this sinogram, which brought us a sinogram with the continuity and a better reconstruction image. The area with high x-ray impermeability is moving inside the sinogram showing a sinusoidal pattern (black arrows). In its reconstructed image, this area is placed in the lower left part (white arrow).

(c) The center of the x-ray impermeable material is set as the fixed point and it is applied to the function $T_{0,\varphi}$; the point is now the line across the center in the sinogram (black arrows). It is placed on the center of the image in the reconstruction (white arrow). It is also noticeable that its reconstruction image is much clearer.

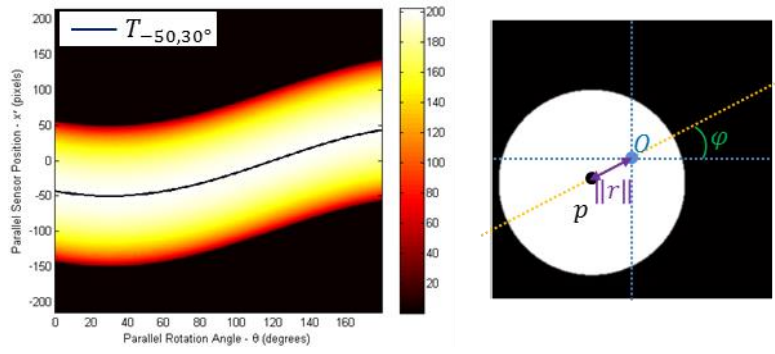

Fig. 4. The sinogram and its reconstruction image with which we figured $\overrightarrow{P_{CA}}$ of each column in Fig. 2(b) sinogram and align them on the function $T_{-50,30°}$.

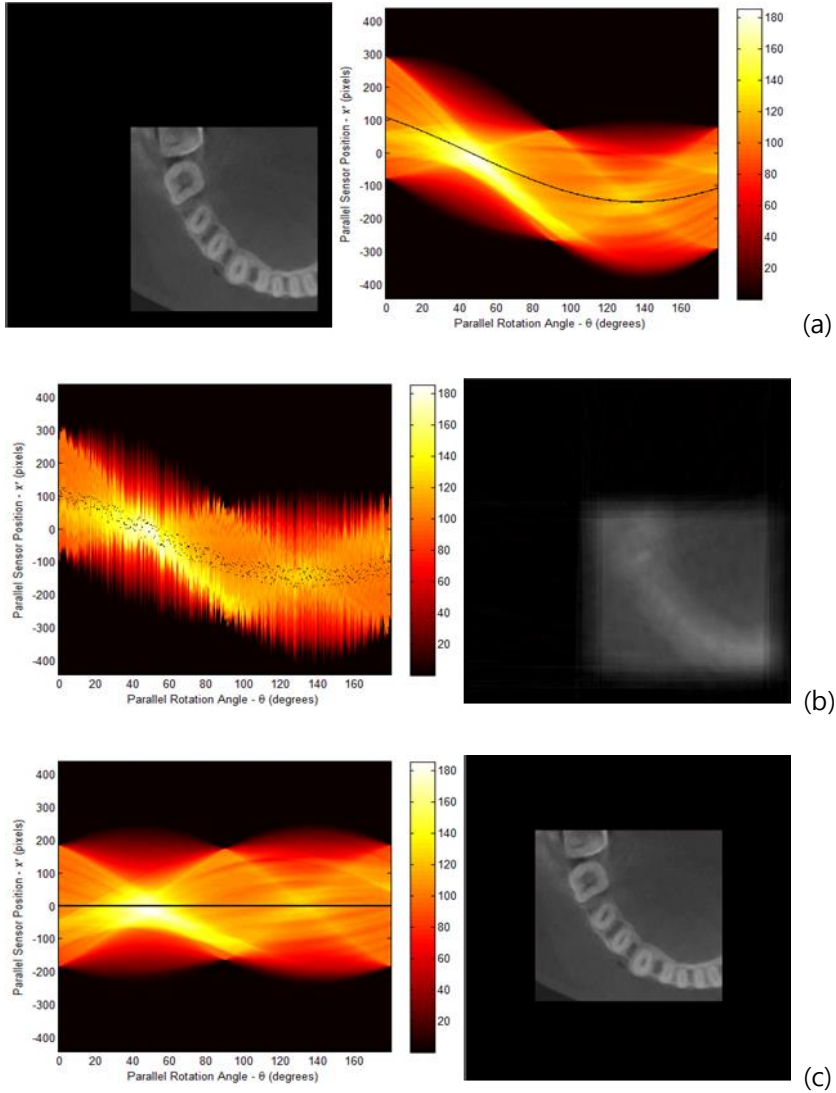

Fig. 5. Analysis of a partial image of human lower jaw including the teeth

(a) An image of the sample and its sinogram (The object is on the upper left side of the stage). The $\overrightarrow{P_{CA}}$s of all projections are marked black in the sinogram. They follow the circular trajectory in the real space, therefore show sinusoidal graph in the sinogram.

(b) A sinogram with translation errors including vertical and horizontal movement at each angle and its reconstruction image. The $\overrightarrow{P_{CA}}$s, the black marks are all scattered.

(c) A sinogram that has $\overrightarrow{P_{CA}}$ aligned on $T_{0,\varphi}$ and its reconstruction image. The $\overrightarrow{P_{CA}}$s are arranged linearly on the center of the sinogram. $\overrightarrow{CA}$ is on the center of the stage and the image is ideally restored.

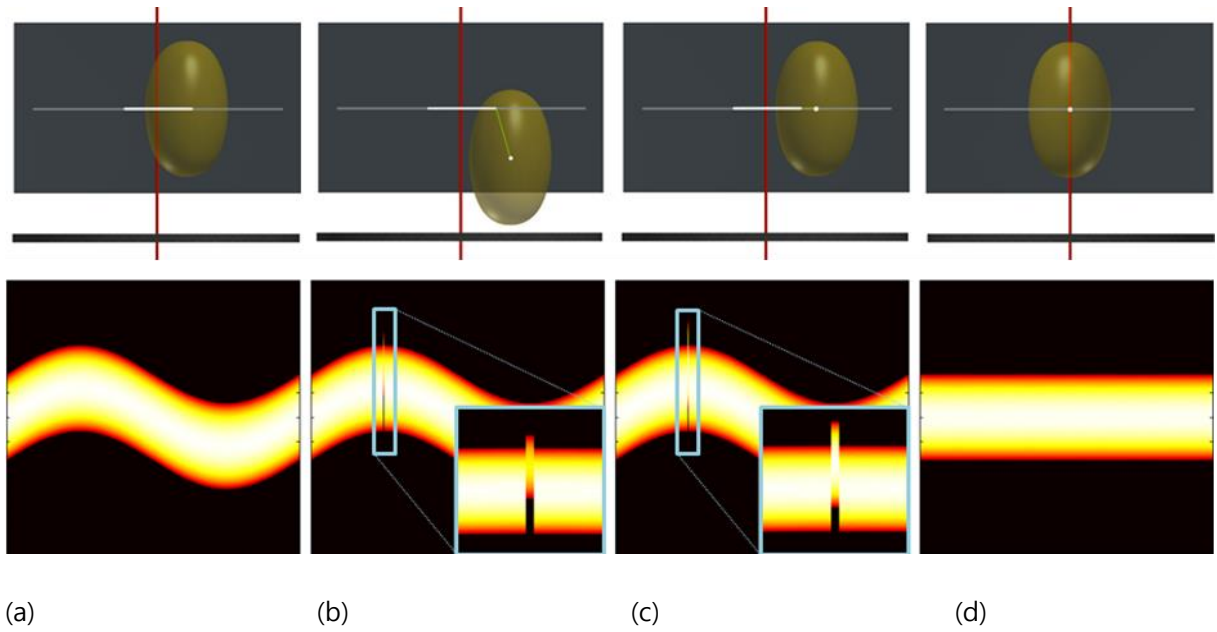

(a)                  (b)                  (c)                  (d)

Fig. 6. An object of prolate spheroid is located on the rotating stage, which has the center of CCD as the RA, and rotates for the 360 degrees. The CA is at the height of $h = 0$, and the distance from the center is $r = d$. The azimuthal angle of the object is 0 degree when the $\theta$ is zero. The white dot and the line represent where $\overrightarrow{P_{CA}}$ lies against each $\theta$.

a. It is the sinogram of the $h = 0$ layer when the object is at $\theta = 90°$ and there is not any known error.

b. It is the sinogram of the $h = 0$ layer when there was a translation error at $\theta = 90°$. The location of the object in the involved layer is moved to the lower left on the stage and its shape is narrower, compared to that of (a). This is reflected in the sinogram; the location of shadow and its width have changed (Look at the magnified part of the sinogram).

c. The $\overrightarrow{P_{CA}}$ in the projection of b was translated vertically to the stage, exactly on the layer in which the $\overrightarrow{P_{CA}}$ without error exists, $h = 0$. The location of shadow has not changed, however, the width now becomes identical with the other $\theta$s in the sinogram.

d. When we aligned $\overrightarrow{P_{CA}}$ on the funcion $T_{0,\varphi,0}$, the $\overrightarrow{P_{CA}}$s were all gathered onto one dot on the center of the stage, and the sinogram became linear.

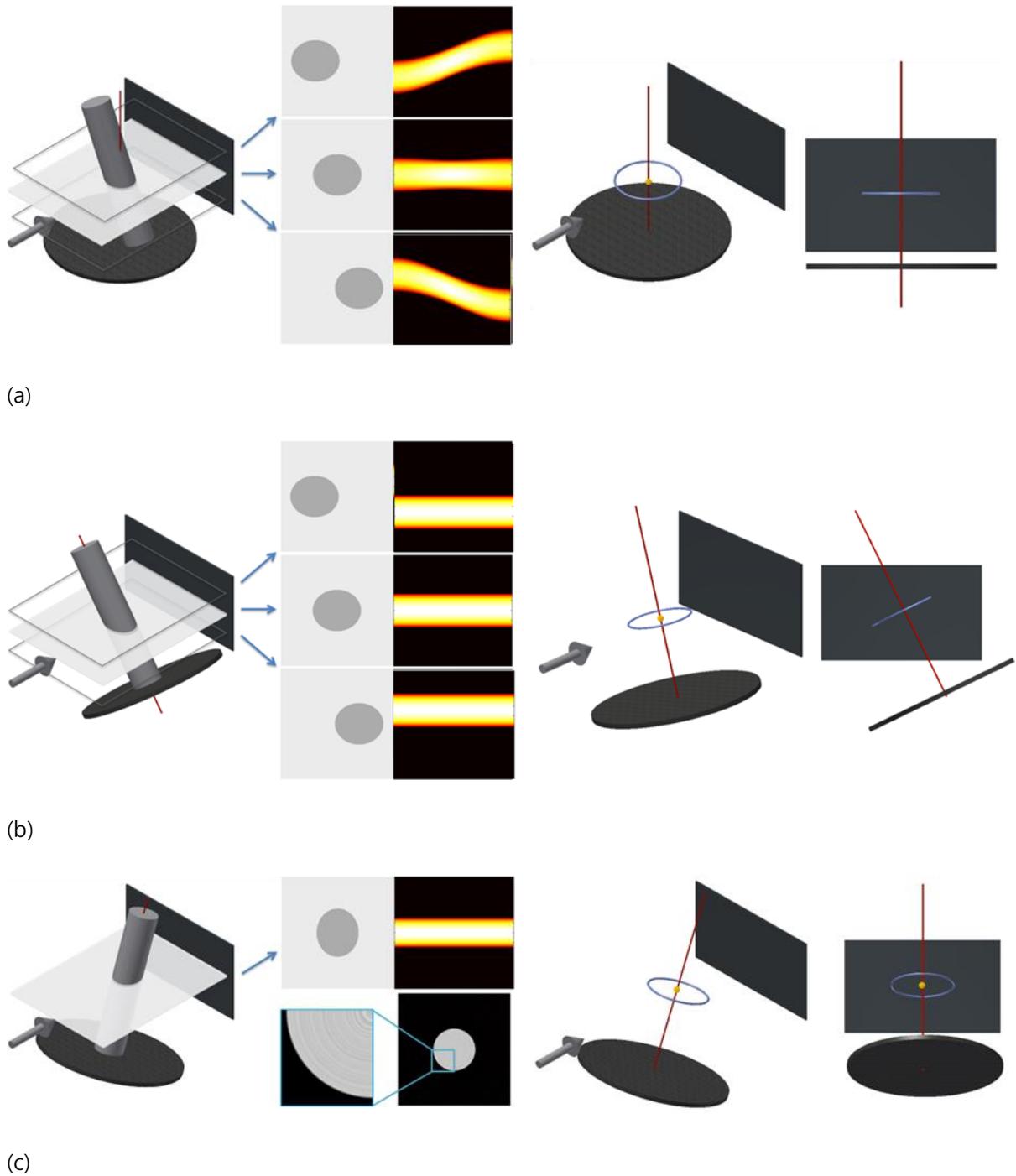

(a)

(b)

(c)

Fig. 7. The sinogram pattern of cylindrical specimen and the $\overrightarrow{CA}$ trajectory in general, depending on the tilt of the object or the RA

(a) The sinogram and the $\overrightarrow{CA}$ trajectory when an object is tilted. Even though the object is tilted, the $\overrightarrow{P_{CA}}$ is on the line parallel with the stage in the projection. This is quite a typical occasion where there is not any error found.

(b) The sinogram and the $\overrightarrow{CA}$ trajectory when the RA is vertically tilted. The object rotates

around the RA and the line that $\overrightarrow{P_{CA}}$ makes are perpendicular to the RA on the projection.

(c) The sinogram, its reconstruction image and the $\overrightarrow{CA}$ trajectory when the RA has a parallel tilt. The pattern of sinogram is linear around the center. However, its reconstructed image is flawed (shown better in the magnified figure), since the layers are all mixed up at each angle θ. In this case, the collection of $\overrightarrow{P_{CA}}$s in the projection makes an oval (elliptical) shape, not a line. The RA is also perpendicular to the major axis of $\overrightarrow{P_{CA}}$ trajectory in this case.

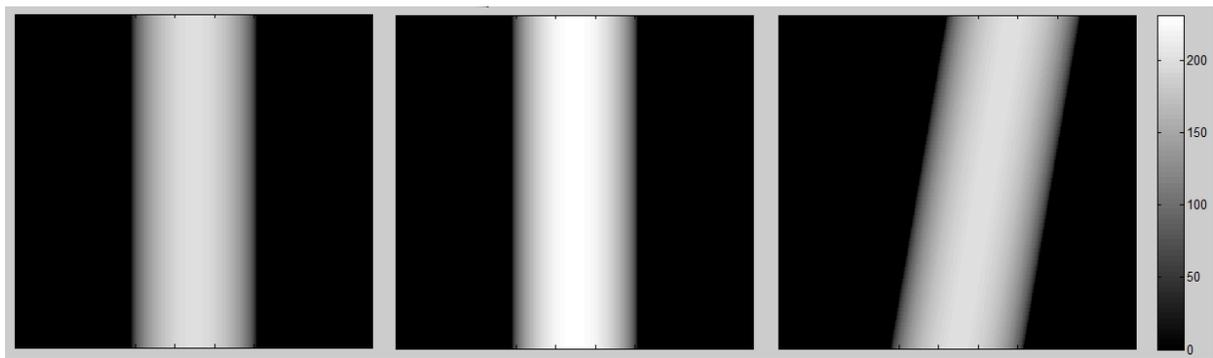

(a) (b) (c)

Fig. 8. The shadow changes depending on the location of a cylindrical specimen

(a) The projection image when the specimen stands upright. When translated without tilts, the shadows are of same shapes even when the locations on the projection are different.

(b) The projection image when the specimen is tilted in parallel with the beam. The shadow is darker than before; the information of the image is changed and the restoration with the projection image is not easy in this case.

(c) The projection image with the vertical tilt of the specimen. The shadow is leaned, however, the information of the image remains the same. In this case, it is possible to restore with the projection image.

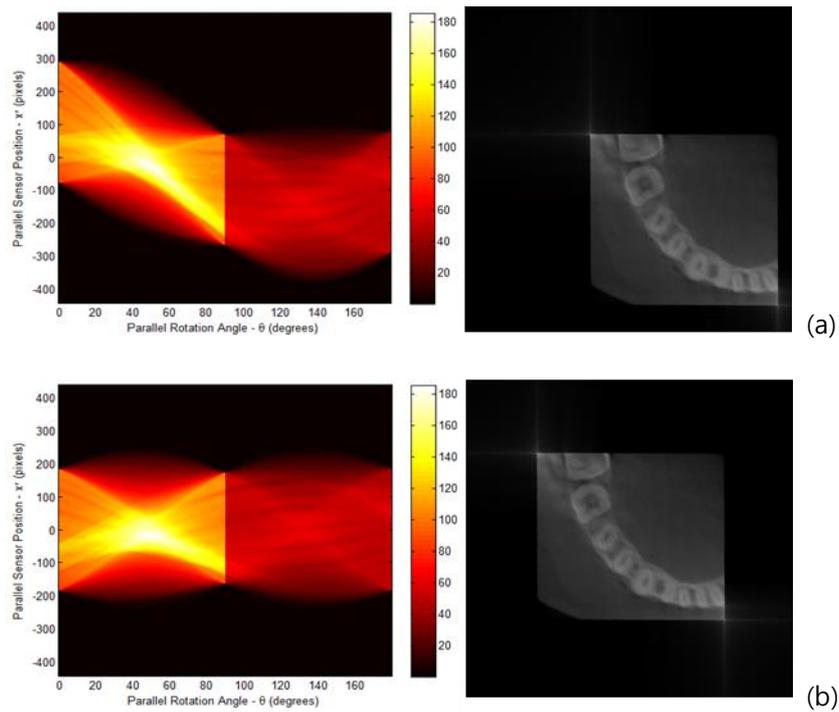

Fig. 9. The cases when the x-ray density of projection is changed during the beam time.

(a) The sinogram and its reconstruction image when the x-ray density of projection in Fig. 5(a) is decreased by half after $\theta = 90°$.

(b) The sinogram and its reconstruction image when the CA was applied to the projection with changed x-ray density.

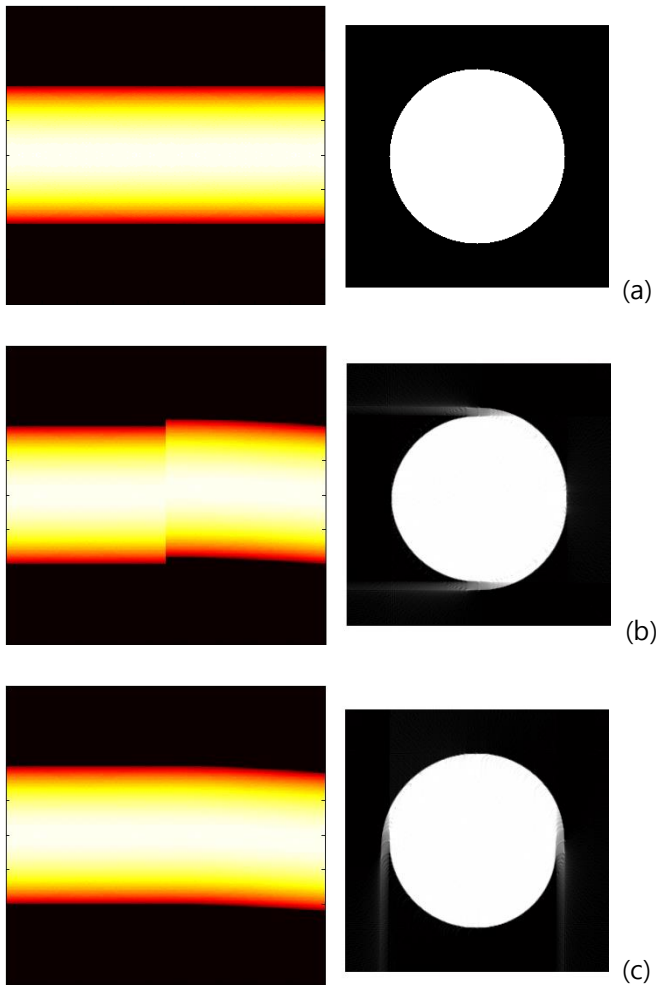

Fig. 10. Vertical Translation with Optimization method

(a) The ideal sinogram and its reconstruction image that were from the specimen of Fig. 1(a)

(b) The sinogram with vertical translation error at $\theta = 90°$ and its reconstruction image. The translation error is found also in the reconstruction image.

(c) The translation error in (b) was optimized meaning that the part of sinogram after the discontinuity was cut and pasted by the distance of translation error in this sinogram. Another error was found in its reconstruction image.